%% file: fornax.tex
\documentclass[fleqn,usenatbib]{mnras}

\usepackage{newtxtext,newtxmath}

\usepackage[T1]{fontenc}
\usepackage{ae,aecompl}

\usepackage{graphicx}	
\usepackage{amsmath}	
\usepackage{amssymb}	
\usepackage{multirow}

\newcommand{\kms}{km~s$^{-1}$}
\DeclareMathOperator{\sech}{sech}

\title[Thick discs in edge-on galaxies]{The imprint of the thick stellar disc in the mid-plane of three early-type edge-on galaxies in the Fornax cluster}

\author[I. Yu. Katkov et al.]{
Ivan Yu. Katkov,$^{1,2}$\thanks{E-mail: katkov@sai.msu.ru (IYK)}
Alexei Yu. Kniazev,$^{3,4,1}$
Anastasia V. Kasparova$^{1}$ and
\newauthor
Olga K. Sil'chenko$^{1,5}$
\\
$^{1}$Sternberg Astronomical Institute, Lomonosov Moscow State University, Universitetsky pr., 13, Moscow, 119234, Russia\\
$^{2}$New York University Abu Dhabi, Saadiyat Island, PO Box 129188, Abu Dhabi, UAE\\
$^{3}$South African Astronomical Observatory, PO Box 9, 7935 Observatory, Cape Town, South Africa\\
$^{4}$Southern African Large Telescope Foundation, PO Box 9, 7935 Observatory, Cape Town, South Africa\\
$^{5}$Isaac Newton Institute, Chile, Moscow Branch
}

\date{Accepted XXX. Received YYY; in original form ZZZ}

\pubyear{2018}

\begin{document}
\label{firstpage}
\pagerange{\pageref{firstpage}--\pageref{lastpage}}
\maketitle

\begin{abstract}
Galactic stellar discs, such as that of the Milky Way, have usually a complex structure consisting of a thin and a thick component.
The study of galactic disc substructures and their differences can shed light on the galaxy assembling processes and their evolution.
However, due to observational difficulties there is a lack of information about the stellar populations of the thick disc components in  external galaxies. Here we investigate three edge-on early-type disc galaxies in the Fornax cluster (IC\,335, NGC\,1380A, NGC\,1381) by using publicly available photometrical data and our new deep long-slit spectroscopy along galactic mid-planes obtained with the 10-m SALT telescope. We report that significant changes of the stellar population properties beyond the radius where photometrical profiles demonstrate a \textit{knee} are caused by an increasing thick disc contribution.
Stellar population properties in the outermost thick-disc dominated regions demonstrate remarkably old ages and a low metallicity.
We interpret these findings as a consequence of star formation quenching in the outermost regions of the discs due to ram pressure gas stripping from the disc periphery at the beginning of the cluster assembly while subsequent star formation occurring in the inner discs being gradually extinguished by starvation.
\end{abstract}




\section{Introduction}
\label{sec:intro}

Thick stellar discs identified as distinct large-scale components of disc galaxies were initially discovered in S0 galaxies through surface photometry of edge-on objects \citep{tsikoudi,burstein79}.
Later \citet{gilmore_reid} found a similar structure in the Milky Way that is a spiral galaxy of a rather late morphological type.
By studying individual stars belonging to the thick disc of the Milky Way, researchers have recognized that the thick disc is an old, rather metal-poor, and magnesium-overabundant component \citep{fuhrmann,Bensby2007ApJ}.
Despite the fact that thick discs are nearly ubiquitous in the local galaxies \citep{comeron,comeron2018} including low-mass dwarf galaxies \citep{Yoachim2006AJ} the spectral studies of thick stellar discs are still rare due to the observational difficulty caused by their low surface brightness.
Only a few recent studies have begun to investigate stellar population properties in the thick discs of various galaxies \citep{YoachimDalcanton2005ApJ,dalcanton,Comeron2015AA_eso533-4,Comeron2016AA_eso243-49,Guerou2016AA_n3115,kasparova} using long-slit spectroscopy with the slits oriented parallel with respect to the galaxy mid-plane or through integral field spectroscopy of edge-on galaxies.

Photometrical studies show that radial surface brightness profiles of galaxies often have breaks with a down-bending (truncation) or an up-bending (antitruncation) shape or a combination of both \citep[][and references therein]{erwin05,Pohlen2006AA,Erwin2008AJ,comeron}.
\citet{comeron}, by studying the photometry of a large sample of edge-on galaxies in the 3.6$\mu$m and 4.5$\mu$m images from the S$^4$G project \citep{s4g}, have concluded that the antitruncated type of galactic disc surface brightness profiles \citep{erwin05} may in some cases be caused by the superposition of a truncated thin disc and a thick disc: at the truncation radius the surface density of the thin disc (or the inner disc, to be more precise) drops sharply. Then, with a cut along the galaxy mid-plane we detect the more extended thick disc in the outer parts of the galaxy.
This is also supported by photometrical measurements which claim that thick galactic discs often have longer radial scale lengths than their thin counterparts \citep{burstein79,pohlen,Yoachim2006AJ,comeron}.
Note that photometrical thin/thick disc decompositions for external galaxies are purely geometric, in contrast to those made for the Milky Way which are often based on the age and the metallicity (or the $\alpha$-enhancement).
Nevertheless this photometrical point of view helps in the interpretation of long-slit spectral data obtained along the mid-planes of edge-on disc galaxies.

In this work we present our new spectral data for three edge-on lenticular galaxies (IC\,335, NGC\,1380A, NGC\,1381) that belong to the Fornax cluster.
These galaxies have been the subject of many spectroscopic studies of their kinematics \citep{DOnofrio1995, ChungBureau2004, Bedregal2006kin, Spolaor2010kin, Vanderbeke2011} as well as stellar populations \citep{Kuntschner2000, Terlevich2002,Bedregal2008stpop, Marmol-Queralto2009, Spolaor2010stpop, koleva, Johnston2012}.
Nevertheless, the above papers considered these objects within galaxy samples and not on an individual basis although the detailed description of their radial profiles (velocities, velocity dispersions, age and metallicities) has been presented, for instance, in \citet{koleva} and \citet{Spolaor2010stpop}.
Our spectral measurements are rather deep, and we have reached the outermost parts of these galaxies and detected strong changes in the stellar population properties at some crucial radii.
Combining spectroscopic evidence with photometrical data we argue that the changes of stellar population properties are caused by the increasing thick disc contribution in the outermost galactic regions along the galactic mid-plane.
We discuss how environment-driven mechanisms in the Fornax cluster could lead to the observed properties of these galaxies.

In Section~\ref{sec:obs}, we describe the observations and the data reduction process including our new framework to correct for the scattered light in the spectra.
In Section~\ref{sec:data_results} we analyse our spectroscopic and archival photometric data and present the results.
A discussion and our main conclusions are provided in Sections~\ref{sec:discussion} and \ref{sec:summary}, respectively.

\section{Observations and data reduction}
\label{sec:obs}

\subsection{Observations}

We have performed long-slit spectroscopy for the following Fornax cluster members: IC~335, NGC~1380A, and NGC~1381.
All of them are lenticular galaxies and are studied in the frame of our project on the stellar populations properties in the early-type disc galaxies in clusters.
Observations were performed with the Robert Stobie Spectrograph \citep[RSS;][]{Burgh03,Kobul03} at the Southern African Large Telescope (SALT) \citep{Buck06,Dono06}.
We have used the long-slit mode with a slit width of 1.25~arcsec and the volume-phase grating GR900 providing a spectral resolution of 4.8~\AA\ (FWHM) in the $3800-6800$~\AA\ spectral range.
All observational details are given in Table~\ref{tbl:Obs}.
The slit was aligned with the galaxies major axes going through their nuclei and mid-plane.
The total exposure time per object is about 1$^{\rm h}$20$^{\rm m}$.
The seeing during the observations was in the {\it $1.5-3.5$}~arcsec range.
The RSS pixel scale is 0.1267~arcsec, and the length of the slit is 8~arcmin.
We used a binning factor of 4 to get a final spatial sampling of 0.507~arcsec pixel$^{-1}$.
An Ar comparison arc spectrum was exposed to calibrate the wavelength scale after each observation.
Spectral flats were taken regularly to correct the spectra for pixel-to-pixel variations.
Spectrophotometric standard stars were observed during twilights, after observing the objects, for a relative flux calibration.

\begin{table}
	\centering
	\caption{Parameters of the long-slit spectroscopy of the studied galaxies.}
	\label{tbl:Obs}
	\begin{tabular}{ccccccc} 
		\hline
		\hline
		Galaxy & Date & Exposure & PA & Seeing\\
		       &      & (sec)    & ($\deg$) & (arcsec) \\
		\hline
		\multirow{2}{*}{IC\,335}    &
                2015-12-09  & 1200$\times$2 &  84   & 2.2--2.7  \\
            &   2016-02-13  & 1280$\times$2 &  84   & 2.7       \\
		\hline
		\multirow{2}{*}{NGC\,1380A} &
                2015-12-16  & 1200$\times$2 &  179  & 3.5 \\
            &   2016-02-14  & 1300$\times$2 &  179  & 2.7 \\
		\hline
		\multirow{2}{*}{NGC\,1381}  &
                2015-12-08  & 1200$\times$2 &  139  & 1.5 \\
            &   2016-01-31  & 1300$\times$2 &  139  & 3.5 \\
		\hline
	\end{tabular}
\end{table}

\subsection{Data reduction}

The primary data reduction was done with the SALT science pipeline \citep{Cr2010}.
After that, bias and gain corrected long-slit data were reduced as described by \citet{Kn08}.
The accuracy of the wavelength calibration as checked by measuring the sky line [O\i]~$\lambda$5577 was about 0.04~\AA.
The observed galaxies are much smaller than 8~arcmin in diameter, so we used the pure night-sky spectra from the slit edges to subtract the sky background.

\subsection{Scattered light}
\label{scat_light_paragraph}

When analysing the reduced spectra we found that our velocity dispersion measurement are overestimated in comparison with those obtained with a better spectral resolution \citep{Bedregal2006kin,koleva}.
We suspected that the reason for such a behavior is the presence of the diffuse scattered light in the spectrograph.
It redistributes light coming from the bright galaxy center to the galaxy outskirts and contributes to the galactic continuum affecting the contrast of the absorption lines.
Neglecting this effect could lead to the systematic biases in the stellar population property determination.
In this section we describe a framework to calculate the scattered light contribution in the spectra along the slit.

We have used the spectrum of a standard star observed with the same spectral setup as the galaxies. The stellar light profiles along the slit can be expressed as:

\begin{multline}
S_{\rm obs}= F_\star\cdot{\rm PSF_{seeing}}*{\rm PSF_{scat}} \approx \\ \approx F_\star\cdot{\rm PSF_{seeing}}*\Big[ \alpha \cdot \delta + f_{\rm scat} \Big],
\end{multline}
where ${\rm PSF_{seeing}}$ is the point spread function caused by atmospheric perturbations, ${\rm PSF_{scat}}$ -- the full PSF of the light scattering in the telescope plus instrument setting, $F_\star$ is a total stellar flux at a given wavelength, $\alpha$ is a scaling factor, $*$ is the convolution operator and $\delta$ is a Dirac delta function.
In the second approximate equality we assume a two components representation for the full scattering function ${\rm PSF_{scat}}$.
This is motivated by necessity to estimate extended part of the scattering function.
The physical meaning of $1-\alpha$ is the fraction of light that is redistributed within the instrument in accordance to the scattering function $f_{\rm scat}$.
Note that one can use a $\delta$ function for the instrumental PSF core since the diffraction PSF is much narrower than the atmospheric one.

\begin{figure}
\centering
\includegraphics[trim=0.5cm 0 0.75cm 0,clip,width=\columnwidth]{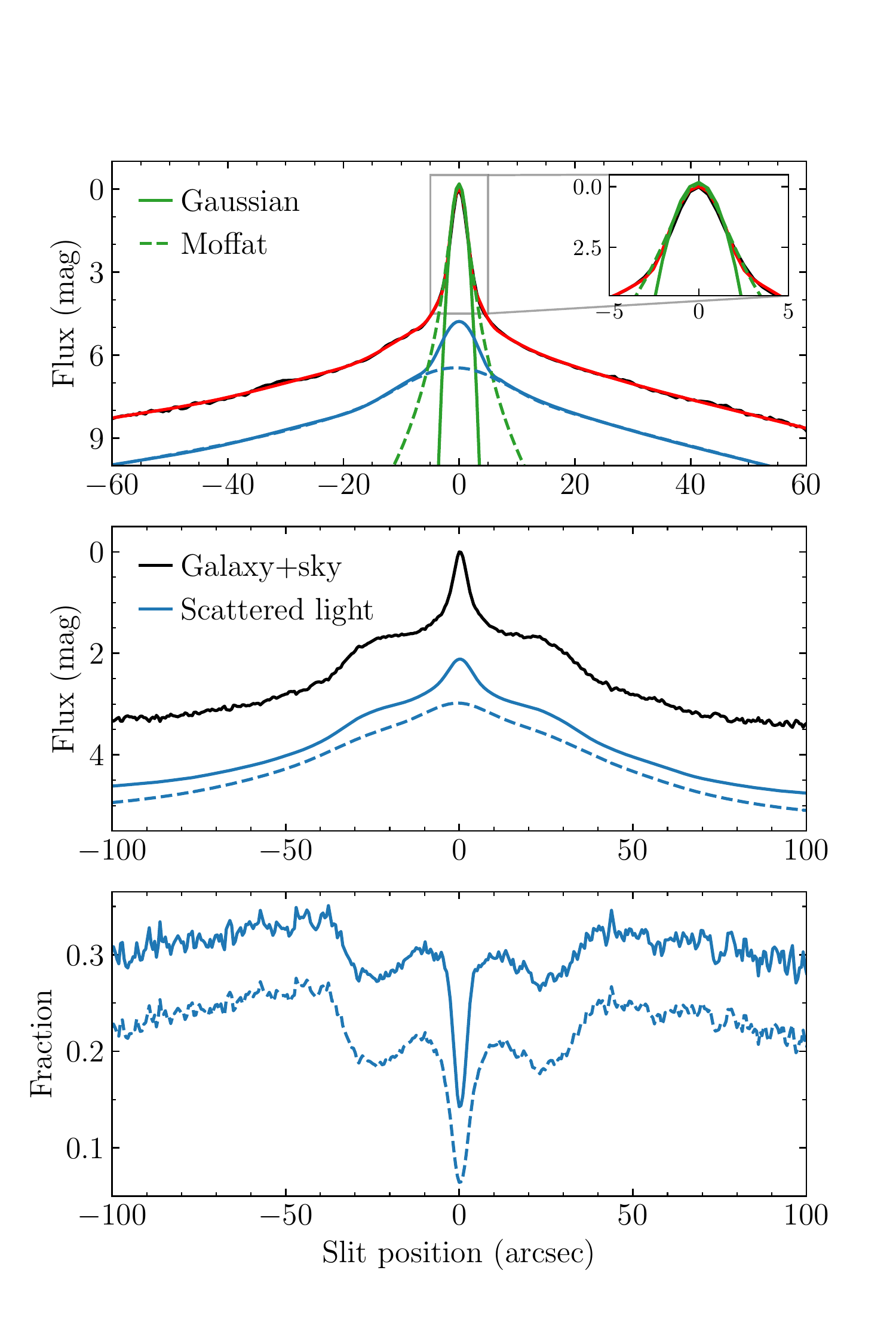}
\caption{
The top panel shows the light profile of a reference star along the slit (black color) which was used for the scattered light component calculation.
The red line represents the result of the convolution of an atmospheric PSF (green lines) with the model of the instrumental PSF component $f_{\rm scat}$ (blue lines).
We used two different of parameterisations for the atmospheric PSF. The Gaussian one is shown with a solid line and the Moffat one with a dashed line.
Both parameterizations provide a different estimation of an additive component of the instrumental scattering function $f_{\rm scat}$.
The middle panel shows the galaxy light profile of NGC\,1380A including the night sky background along the slit at a wavelength 5096 \AA.
The estimated scattered light component, which corresponds to the term in square brackets of equation~(\ref{with_squared_brackets}), is shown by the blue lines corresponding to both Gaussian and Moffat shapes of the atmospheric PSF.
The lower panel indicates the fraction of light scattered by the telescope and instrument with respect to the total light distribution.
See section~\ref{scat_light_paragraph} for a detailed description. \label{scat_light_fig}}
\end{figure}

This additive representation of the instrumental PSF allows us to propose a procedure for the calculation of $f_{\rm scat}$ in any spectrum.
The top panel of Figure~\ref{scat_light_fig} demonstrates the procedure.

For further calculations we approximated the shape of $F_\star\cdot{\rm PSF_{seeing}}$ either by a Gaussian or a Moffat profile fitting the upper part ($>I_{\rm max}/3$, where $I_{\rm max}$ is the maximum of the light profile) of the observed profile $S_{\rm obs}$ and scaled to contain the full stellar flux along the slit.
We are not able to recover the wings of the $F_\star\cdot{\rm PSF_{seeing}}$ profile because the profile wings are affected by the instrumental light scattering.
So we used these two extreme parametrisations (Gaussian and Moffat) having completely different wings and considered how they affect the final parameters.

To avoid degeneracies in the main fitting procedure we fixed the scaling coefficient $\alpha$ by requiring equality between the maximum values of the Gaussian/Moffat representation (green line in Fig.~\ref{scat_light_fig}) and the observed stellar profile (black line).
In the main minimization loop we approximate the $f_{\rm scat}$ function as the sum of three exponential functions and three Gaussians (blue line in Fig.~\ref{scat_light_fig}) which results in a good modeling of the observed stellar profile.
To take into account variations along the wavelengths we determined $f_{\rm scat}$ in 6 bins along the whole spectral range from 3800 till 6800~\AA.
The shapes of $f_{\rm scat}$ for different wavelength bins turned out to be very similar.
The total fraction of the scattered light not accounted by the atmospheric ${\rm PSF_{seeing}}$ is about 15\,(25)\,per cent for the Moffat (Gaussian) parametrisation.

One can apply the same additive parametrisation of light scattering to a galaxy profile to estimate the contribution of scattered light in the observed spectrum:

\begin{multline}
G_{\rm obs} \approx G_{\rm PSF}*\Big[ \alpha \cdot \delta + f_{\rm scat} \Big] = \\
= \alpha \cdot G_{\rm PSF} + G_{\rm PSF} * f_{\rm scat},
\label{galaxy_profile_expr}
\end{multline}
where $G_{\rm PSF}$ is a galaxy light profile at some wavelength affected by atmospheric seeing only. Convolving $G_{\rm obs}$ with $f_{\rm scat}$ one and then two times we obtain equations (\ref{eq3}) and (\ref{eq4}), respectively.

\begin{equation}
G_{\rm obs} * f_{\rm scat} = \alpha \cdot G_{\rm PSF} * f_{\rm scat} + G_{\rm PSF} * \big(f_{\rm scat} * f_{\rm scat} \big),
\label{eq3}
\end{equation}

\begin{multline}
G_{\rm obs} * \big(f_{\rm scat} * f_{\rm scat} \big) = \alpha \cdot G_{\rm PSF} * \big(f_{\rm scat} * f_{\rm scat} \big) + \\
+ G_{\rm PSF} * \big(f_{\rm scat} * f_{\rm scat} * f_{\rm scat} \big).
\label{eq4}
\end{multline}
Neglecting the last term in the equation~(\ref{eq4})\footnote{This term corresponds to the third order effects in our formalism. The estimated fraction of the scattered light 15(25) per cent for the Moffat(Gaussian) parametrisation, therefore the last term would only have an effect of the order of $0.15^3\approx0.3$ ($0.25^3\approx1.6$) per cent in flux.} and substituting it in the equation~(\ref{galaxy_profile_expr}) one can write:

\begin{multline}
\alpha G_{\rm PSF} = G_{\rm obs} - \Big[ \frac{1}{\alpha} G_{\rm obs} * f_{\rm scat} - \frac{1}{\alpha^2} G_{\rm obs} * f_{\rm scat} * f_{\rm scat} \Big].
\label{with_squared_brackets}
\end{multline}
The term in the square brackets can be considered as an additive component caused by the light scattering within the instrument.

Under this framework we calculated an additive component of the scattered light at each wavelength.
Nevertheless, photons are also scattered along the dispersion direction resulting in a shallowing of spectral absorption features of the scattered light component.
To take into account that fact we convolved the computed component of scattered light along the dispersion direction with the normalised function $f_{\rm scat}$ assuming that the light scattering in the dispersion direction is qualitatively similar to that in the spatial direction\footnote{Note that one can use convolution with a broad enough Gaussian function. We tested this and found completely similar resulting stellar population properties.}.
This step does not affect the flux level but makes absorption features of the scattered component shallower.
To obtain spectra unaffected by instrumental light scattering we subtracted the scattered additive component at every wavelength in our observed spectra.

The middle panel of Figure~\ref{scat_light_fig} shows an application of this framework to the galaxy NGC~1380A.
The bottom panel demonstrates the relative contribution of the scattered light to the galaxy profile for two parametrisations of the atmospheric PSF (Gaussian/Moffat).
Despite our estimation of a total scattered light fraction of 15 (25)\,per cent based on a standard star, the relative contribution of the scattered light at a given position on the slit can reach higher values.
This happens because the PSF is very broad and covers a large fraction of the galaxy.

\section{Data analysis and results}
\label{sec:data_results}

\subsection{Spectroscopy}
\label{sec:data_results_stpop}

Stellar kinematics and stellar population properties resolved along the slit were derived by the full spectral fitting code {\sc NBursts} from \citet{nburst_a,nburst_b}.
This technique implements a pixel-to-pixel $\chi^2$ minimization fitting algorithm, where an observed spectrum is approximated by a stellar population model broadened with a parametric line-of-sight velocity distribution (LOSVD).
We used a grid of PEGASE.HR high-resolution simple stellar population (SSP) models \citep{LeBorgne+04} based on the ELODIE3.1 empirical stellar library \citep{Prugniel07} with a fixed Salpeter initial mass function (IMF) and pre-convolved with the RSS spectrograph instrumental function recovered from the spectrum of a Lick standard star.
The minimisation loop chooses a SSP model by interpolating the age and metallicity values in the grid, then model is broadened with stellar LOSVD and multiplied by a Legendre polynomial to match the continuum shape so to take into account possible internal dust reddening and/or spectral calibration errors both in the data and models.

The used stellar population models are computed for the solar element abundance ratio only because they are based on an empirical library of stars from the solar vicinity.
To check the relative $\alpha$-element abundance we calculated the Lick indices Mgb and $\langle \mbox{Fe} \rangle \equiv \mbox{Fe5270} + \mbox{Fe5335}$ \citep{woretal, worott} and compared them to the SSP evolutionary synthesis models by \citet{thomod}.

To obtain reliable radial profiles we made a spatial binning of the spectra along the slit.
We used linearly increasing bins from 2 pixels at the galaxy center up to 20 pixels ($0.5-10$~arcsec) at the radius where the signal-to-noise ratio (SNR) per unbinned pixel is equal to 3 adjusting the SNR within the bin to be greater than 15 per pixel at $4640\pm10$~\AA.
For the [Mg/Fe] profiles we used a different spatial binning requiring the minimal SNR to be 20 per bin because of the Lick index measurements being more sensitive to the noisy data than the full spectra fitting.

To determine the full spectral fitting parameter uncertainties we carried out Monte Carlo simulation for each spatial bin.
We generated a hundred realization of synthetic spectra by adding a random noise to the best-fitting model corresponding to the signal-to-noise ratio in the bin.
Then we fitted each synthetic spectrum and estimated the errors as the standard deviation of the output model parameters.
The uncertainties for the Mgb and $\langle \mbox{Fe} \rangle$ indices were computed by using photon Poisson errors propagated through all the data reduction steps.

Having solved the scattered light problem we compared our velocity dispersion measurements with the data obtained with the ESO FORS2 spectrograph with significantly higher resolution ($\sigma_{\rm inst}\approx20$ \kms) by \citet{Bedregal2006kin}.
We found that the use of Gaussian shape for the atmospheric ${\rm PSF_{seeing}}$ provides velocity dispersion measurements in good agreement with the higher resolution data down to $\sigma_{\rm inst}/2 \approx 60$~\kms. The instrumental resolution is $\sigma_{\rm inst}\approx 120$~\kms\ around Mgb band at 5100\AA.
This is sufficient for measuring  the velocities, however, the measurements of the velocity dispersions in the cold thin component of stellar discs ($\sigma_\star$ as low as $20-40$~\kms) could be affected by a systematic bias to higher values.
At the same time the subtraction of the scattered light component computed with the Moffat parametrisation of ${\rm PSF_{seeing}}$ provides velocity dispersions overestimated by 20~\kms\ as well as an underestimation of stellar metallicities by 0.1~dex at the 60~\kms\ level of velocity dispersion compared to \citet{Bedregal2006kin}.
For higher velocity dispersions the biases become negligible.
So through the following stellar population analysis we used only spectra with removed scattered light component computed with the Gaussian shape of ${\rm PSF_{seeing}}$.

\begin{figure*}
\centerline{
    \includegraphics[clip,trim={0.0cm 0 0.2cm 0},height=0.85\textheight]{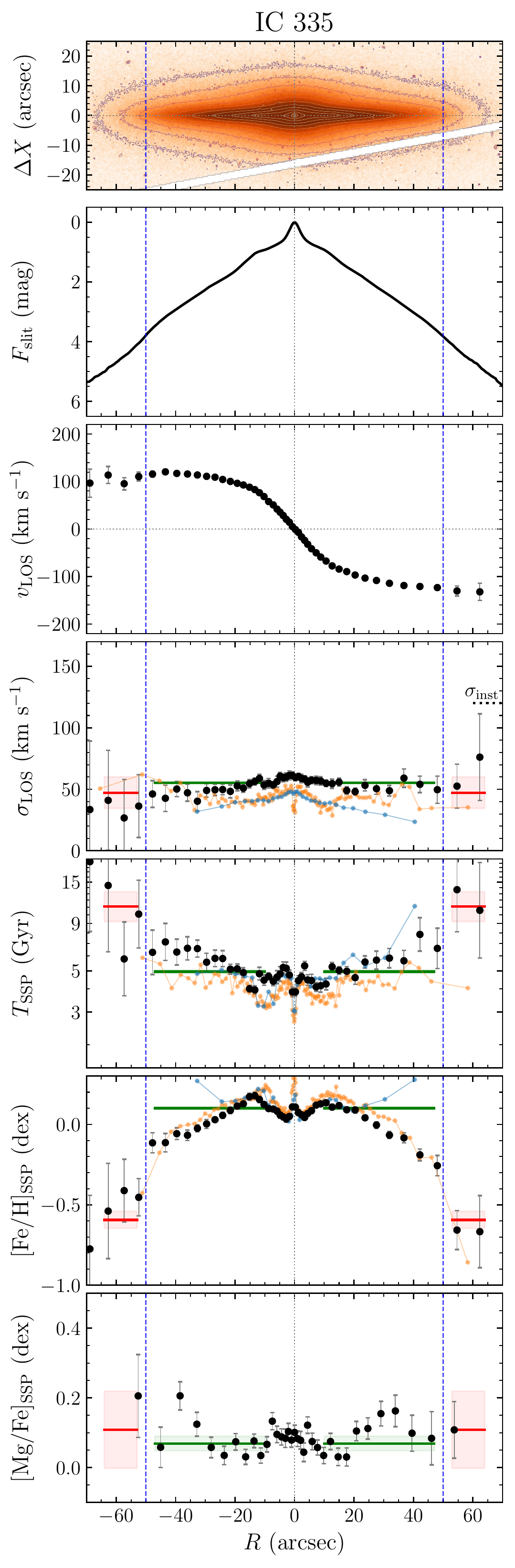}
    \includegraphics[clip,trim={1.8cm 0 0.2cm 0},height=0.85\textheight]{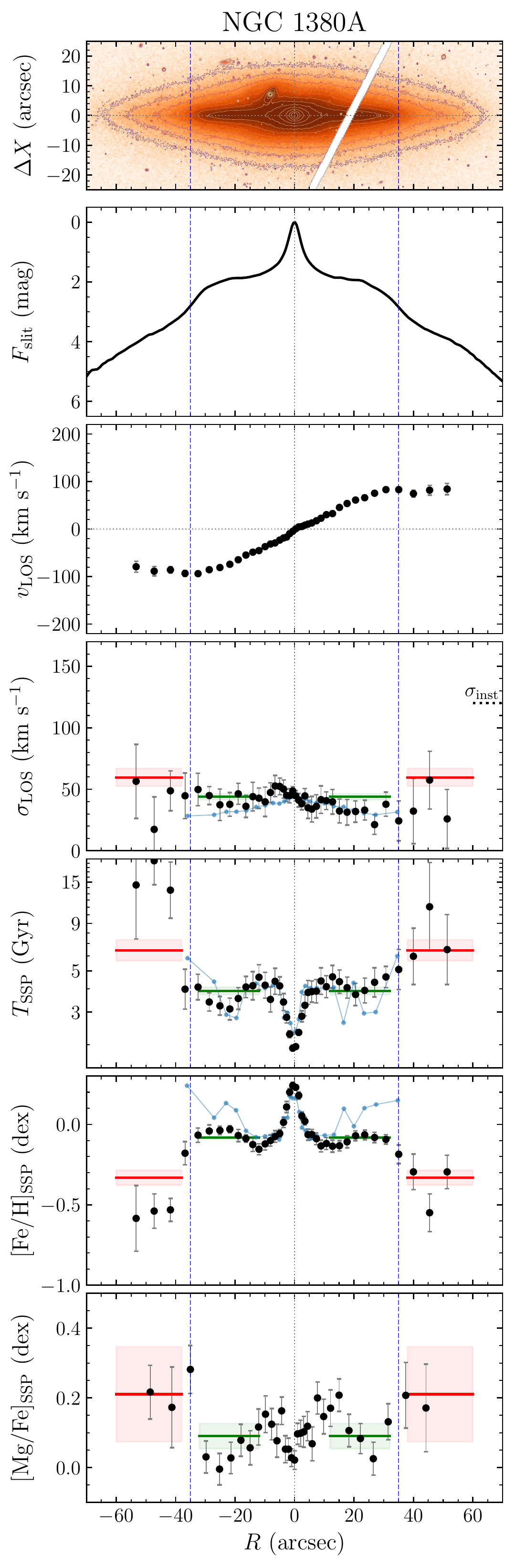}
    \includegraphics[clip,trim={1.8cm 0 0.2cm 0},height=0.85\textheight]{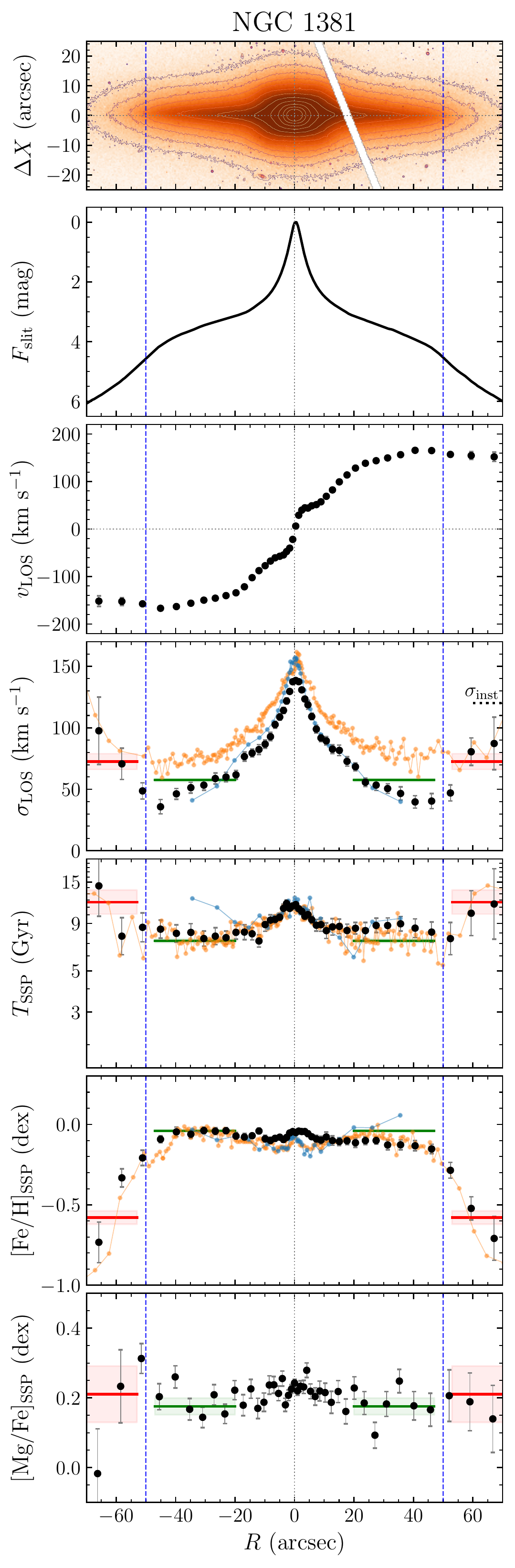}
    }
\caption{Radial profiles of stellar population properties recovered from the long-slit SALT spectra by the means of the full spectral fitting.
The top panels show reference images which are also used in the photometrical analysis (see Section~\ref{sec:data_results_photometry}).
Surface brightness profiles of galaxies extracted from long-slit spectra are shown in the second line panels.
These profiles are normalized so that the center has a zero magnitude.
The next two rows demonstrate line-of-sight stellar velocities and velocity dispersions.
The remaining two rows correspond to the SSP equivalent measurements of ages, metallicities and magnesium-to-iron ratios [Mg/Fe].
[Mg/Fe] profiles were derived with Lick indices measurements, which required higher signal-to-noise ratio than the SSP fitting and a different spatial binning.
Green and red horizontal lines demonstrate measurements recovered from the binned spectra and corresponding uncertainties are shown in the shaded area around these lines.
The profiles coloured in orange and blue with connected small dots are taken from \citet{koleva}.
Orange symbols correspond to a GMOS dataset with a comparable spectral resolution to our RSS data, then blue ones show FORS2 data with a six times better resolution ($\sigma_{\rm inst}\approx20$ \kms).
Vertical blue dotted lines correspond to radii where photometrical profiles indicate a \textit{knee}.}
\label{fig_specfit}
\end{figure*}

We present the profiles of the SSP-equivalent stellar ages and metallicities and stellar kinematics obtained by full spectrum fitting in Fig.~\ref{fig_specfit}.
The main feature of all studied galaxies is a strong change in the properties of the stellar populations, in particular in the stellar metallicities, beyond a certain radius.

Our photometrical analysis (see Section~\ref{sec:data_results_photometry} for details) revealed that the studied galaxies have complex disc structure.
We denoted the \textit{knee} radii where a break (truncation) in the surface brightness profile as well as significant changes in the stellar population parameters have appeared.
The \textit{knee} separates the inner and the outer discs.
To compare their properties we binned the spectra in two regions separated by the \textit{knee} radius correcting for the LOS velocity variations, and analyzed these in the same manner as the original radial bins.
We used luminosity weighted integration within the bins $R_{\rm b}<R<R_{\rm knee}-3^{\prime \prime}$ and $R>R_{\rm knee}+3^{\prime \prime}$ for the inner and outer discs correspondingly (see Table~\ref{tab:knee_tab}).
To avoid possible contributions from the bulge and/or the bar we exclude the central disc regions ($R<R_{\rm b}$) choosing $R_{\rm b}$ by eye.
We defined the maximum radii for the outer radial bins as the radii where $S/N=3$ per unbinned pixel.
The properties of the inner and the outer disc are presented in Table~\ref{tab:stpop_binned} and are shown in Fig.~\ref{fig_specfit} via green and red lines.

\begin{table}
\begin{center}
\caption{Approximate radii of the bulges/bars and knees separating the different segments of the brightness profiles.}
\begin{tabular}{lccc}
\hline
\hline
                   & IC\,335 & NGC\,1380A & NGC\,1381 \\
\hline
$R_{\rm b}$ (arcsec)    & 10      & 12         & 20  \\
$R_{\rm knee}$ (arcsec) & 50$\pm$3      & 35$\pm$3         & 50$\pm$3  \\
\hline
\end{tabular}
\label{tab:knee_tab}
\end{center}
\end{table}

The detailed stellar population profiles of IC\,335, NGC\,1380A, NGC\,1381 were already obtained with ESO/FORS2 by \citet{Bedregal2006kin,Bedregal2008stpop} and with Gemini/GMOS data by \citet{Spolaor2010stpop} and later re-analysed by \citet{koleva} by using a full spectral fitting technique.
We compare our profiles with those by \citet{koleva} in Fig.~\ref{fig_specfit} (blue and orange lines with dots) and found that our measurements are in good agreement with theirs.
Below we comment on the kinematics and the stellar population profiles of the galaxies in our sample.

\textbf{IC\,335} (IC~1963, FCC~153): This galaxy demonstrates quite a flat velocity dispersion profile ($\sigma_\star\approx60$ \kms) which is in agreement with the photometrical decompositions where the bulge was fitted by a point-like source \citep{s4gdecomp} or as a S{\'e}rsic function with a very small effective radius \citep{comeron2018}. In both cases the bulge contains only $2-6$~\% of galaxy light.

The stellar age profile demonstrates a steady increase of age from $T_{\rm SSP}\approx5$~Gyr at the central region of the disc to $>10$~Gyr at the outskirts.
The metallicity profile has smooth variations with local maxima at $|R|=10...15$~arcsec and gradually decreases to a value of [Fe/H]$_{\rm SSP}=-0.1$ dex at $\approx40$~arcsec beyond which it drops further down to [Fe/H]$_{\rm SSP}\approx-0.5$~dex.

Generally, the $\alpha$-elements ratio is slightly positive ([Mg/Fe]$\approx0.1$~dex) over the inner disc with larger values at radii larger than 25~arcsec.
The analysis of the spectra binned over the whole inner and outer discs does not show significant differences in the $\alpha$-elements abundances.

\textbf{NGC\,1380A} (FCC~177): This galaxy also has a small bulge component as indicated by the flat velocity dispersion profile.
The central region of the galaxy (inner $\pm5$~arcsec) shows signatures of rejuvenation of the stellar population due to a recent star formation event: while the inner disc has an average ages of $4$~Gyr and a metallicity of $-0.1$~dex, the central region has $T_{\rm SSP}=2$~Gyr and [Fe/H]$_{\rm SSP}=+0.2$~dex.
Again we detected a significant decrease of the stellar metallicity at the \textit{knee} radius ($R\approx35\arcsec$) down to a value of [Fe/H]$_{\rm SSP}=-0.5$~dex.
The averaged ages and metallicities for the outer disc are not as different from those of the main disc and one could suppose just by looking at the individual data points.
This is because the luminosity-averaging causes the innermost points to dominate.
The [Mg/Fe] profile has a large scatter but it seems that the last detected points have slightly higher values than the regions at $|R|=20\dots30$~arcsec.

\textbf{NGC\,1381} (FCC~170): This is the only galaxy in our sample with a prominent bulge that can be clearly seen on the galaxy image (top right panel in Fig.~\ref{fig_specfit}) as well as on the stellar population profiles.
Stellar kinematics show double-humped features on the line-of-sight velocity profile and a shoulder in the velocity dispersion profile that are in agreement with the measurements by \citet{ChungBureau2004} and \citet{Bedregal2006kin} and indicate the presence of a bar \citep{BureauAthanasoula2005}.
The metallicity profile looks flat within the inner disc.
Beyond $R=50$~arcsec the metallicity decreases down to [Fe/H]$_{\rm SSP}\approx-0.5\dotsc-0.7$~dex.

The age profile stays constant within the inner disc at level of $8-9$~Gyr and increases to $10-11$~Gyr in the outskirts.
The [Mg/Fe] element ratio indicates that stellar populations are enriched by $\alpha$-elements throughout the entire disc ([Mg/Fe]~$\approx0.2$~dex).

\input{tbl_stpop_binned.tex}

\subsection{Photometry}
\label{sec:data_results_photometry}

\begin{figure}
   \includegraphics[clip,trim={0 0.8cm 0 0},width=0.47\textwidth]{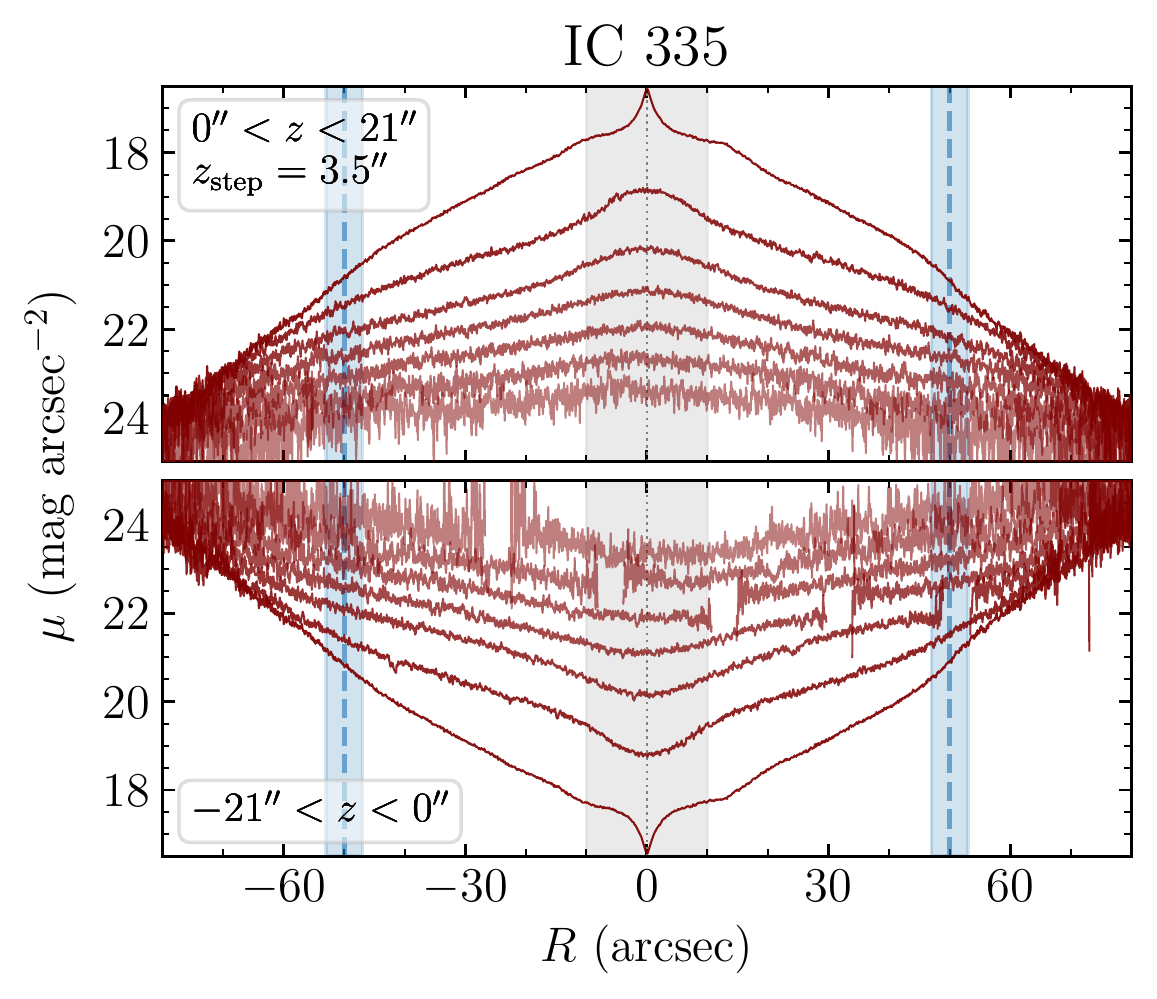}\\
   \includegraphics[clip,trim={0 0.8cm 0 0},width=0.47\textwidth]{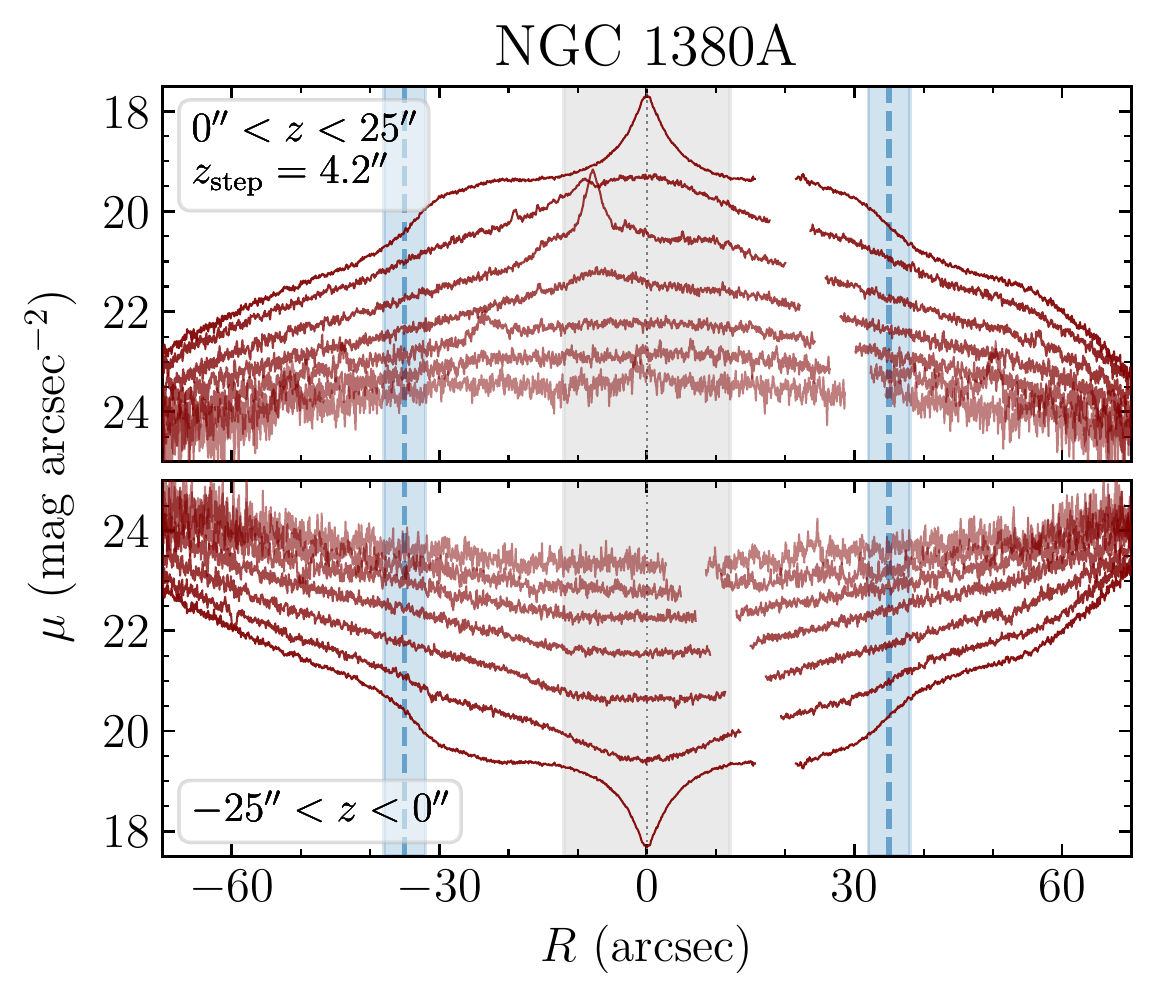}\\
   \includegraphics[width=0.47\textwidth]{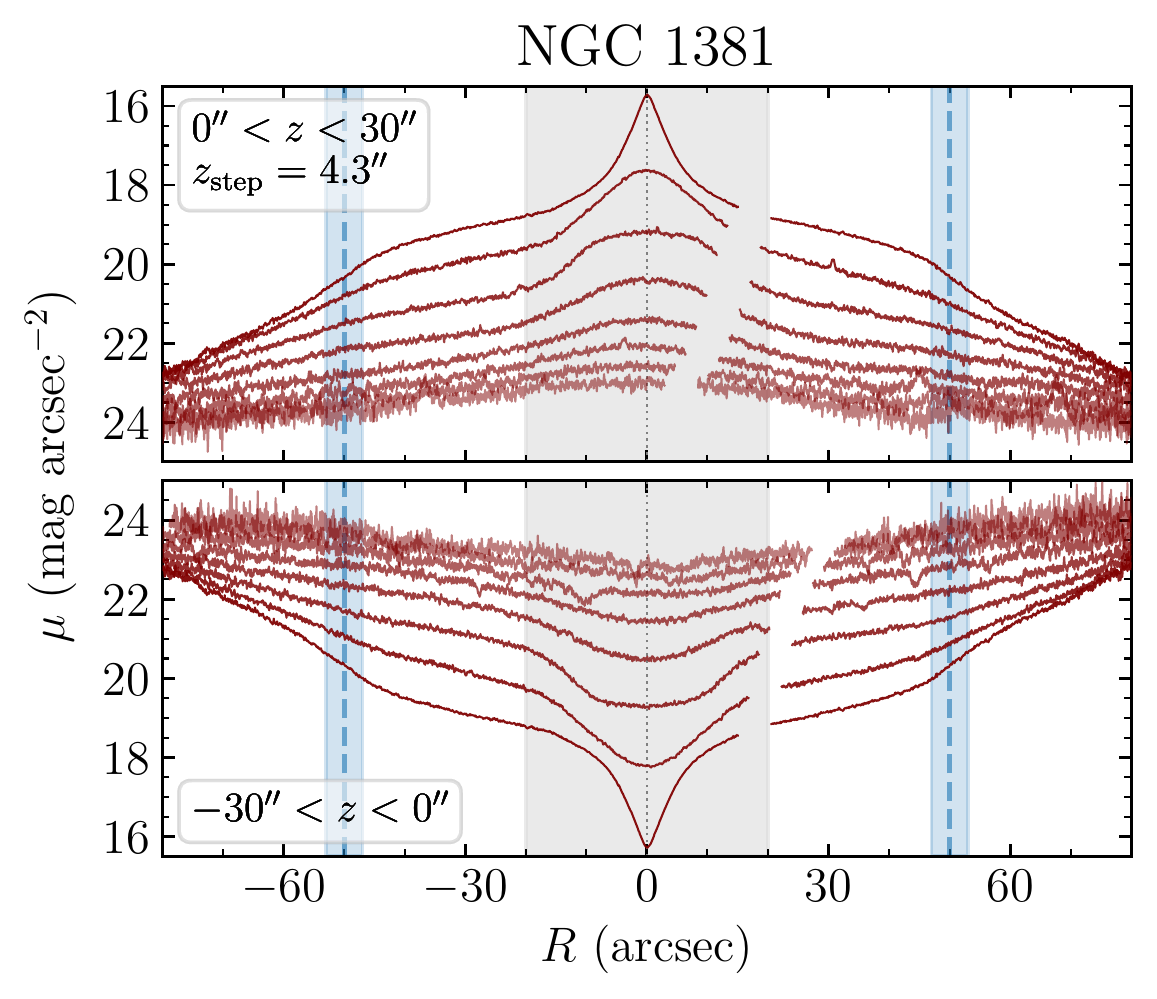}
\caption{The radial photometric profiles at different $z$-distances below and above the main planes of the galaxies. The step between cuts parallel to the mid-plane ($z$-step) is shown in each panel. The central shaded regions are excluded from the analysis while blue stripes show the \textit{knee} radii.}
\label{phot_prof}
\end{figure}

Since our spectral analysis has shown sharp changes of the stellar population parameters at the \textit{knee} radius, we would like to understand whether this happens due to the intrinsic features (internal gradients) of the thin stellar disc or due to changes in the relative contribution of the thin and thick disc subsystems in the radial direction.
Are there any reasons from the photometric point of view to assume that thick disc stars in the studied galaxies have a considerable contribution to the  light in the mid-plane at radii $R \gtrsim 30 - 50$~arcsec?

\subsubsection{Radial structure}

We have used HST images of IC\,335, NGC\,1380A and NGC\,1381 obtained with ACS/WFC in the F850LP band.
In Fig.~\ref{phot_prof} we present the radial photometric profiles at different $z$-distances below and above the main plane of the galaxies.
Radial profiles have a broken profile structure, and the \textit{knees}, separating the different exponential sections, separate also the regions with different stellar populations (see Fig.~\ref{fig_specfit}).
The knee radii of each galaxy are listed in Table~\ref{tab:knee_tab}.
It is well seen (Fig.~\ref{phot_prof}) that the knees at higher galactic altitudes are less prominent, therefore the radial profiles outside the main plane ($z\gtrsim10$~arcsec) can be described by a single exponential law.
\citet{Martinez-Lombilla2018} found a similar behavior for two highly inclined nearby galaxies NGC~4565 and NGC~5907.

\begin{figure*}
\centerline{
    \includegraphics[width=0.99\textwidth]{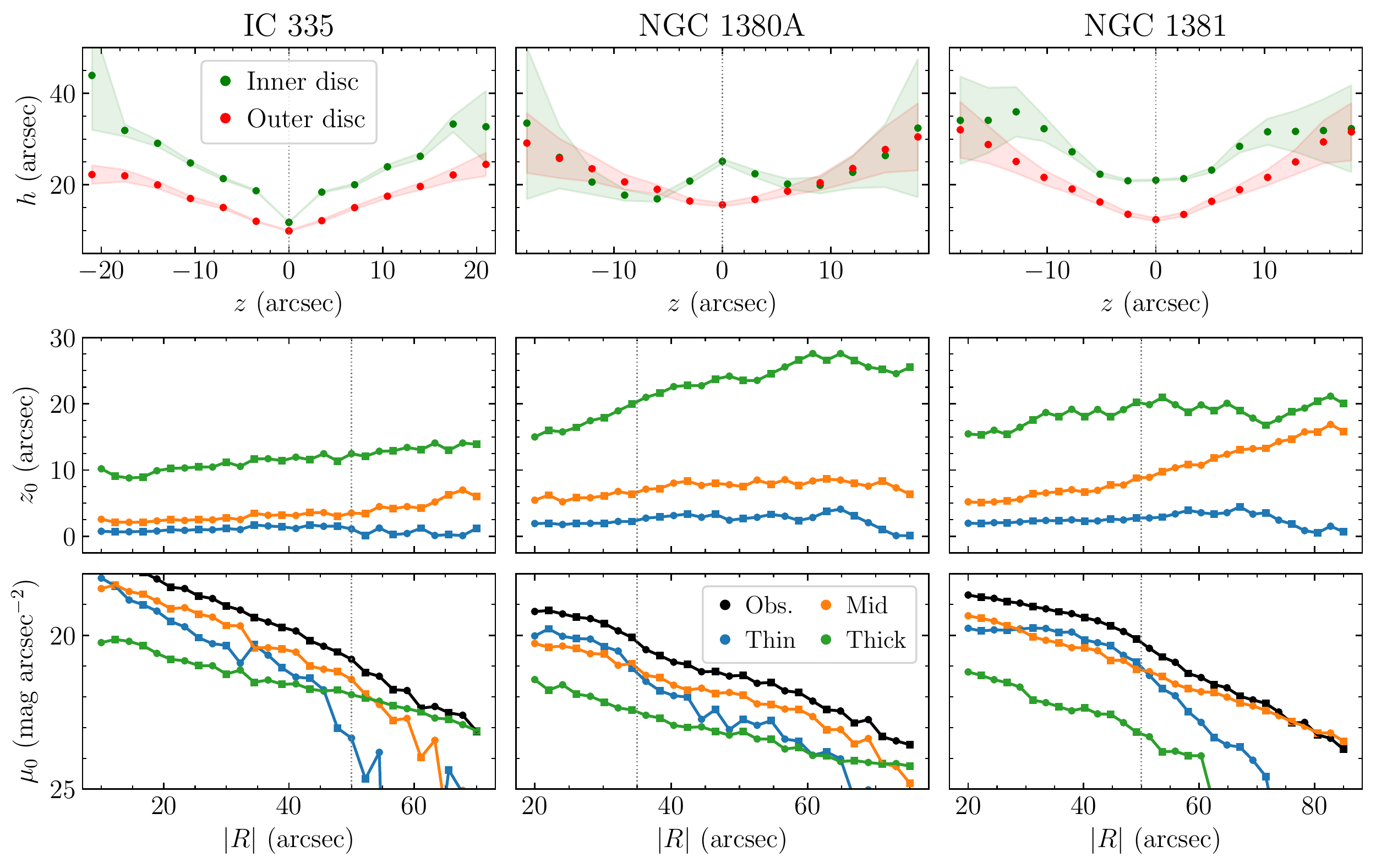}
}
\caption{{\it Top panels:} The scale-length $h$ of the outer and the inner segments (red and green lines) as a function of the height, $z$.
{\it Mid panels:}  Blue, orange, and green lines show the scale-height $z_0$ of thin, mid, and thick disc component as a function of $|R|$.
Circles and squared symbols correspond to positive and negative values of $R$, respectively.
{\it Bottom panels: } Mid-plane surface brightnesses in the ACS/WFC F850LP filter for the thin, mid and thick components and observed values of $\mu_0$. The dashed vertical lines show the knee radii $R_{\rm knee}$.
Beyond $R_{\rm knee}$ the contribution of the thinnest disc component  significantly decreases.}
\label{phot_decomp}
\end{figure*}

It is difficult to disentangle correctly the individual contributions of the two discs due to parameter degeneracy and the unknown truncation law.
For this reason, it makes sense to fit each segment of the profile using one exponential component. In this way we get a~radial scale of some superposition of a thick and a thin disc.
We have estimated the radial scale-length inside and outside $R_{\rm knee}$ using the expression \citep{vdKruit1981}
\begin{equation}
I_R(R,z)\propto\frac{R}{h(z)}K_1\left(\frac{R}{h(z)}\right),
\label{eq_exp_edge-on}
\end{equation}
where $h(z)$ is the exponential scale length at a given $z$, and $K_1$ is the modified Bessel function.
In order to analyse only the data concerning the disc components, we exclude the inner regions of the galaxies $R<R_{\rm b}$, where the influence of spherical subsystems or bars is possible.

In Fig.~\ref{phot_decomp}, on the top row, we present the radial scale length of the outer and inner segments (red and green signs) as a function of the $z$-distance from the mid-plane.
We see a clear trend for the radial scale length to grow with increasing $z$, which can be interpreted as an evidence of the disc heterogeneity.
For two of the three galaxies~-- NGC\,1380A and NGC\,1381~-- we see that there are significant differences in the radial scale lengths of both segments only within the inner layer with $|z|<6-12$~arcsec.
This corresponds to the fact that the broken profile becomes a single exponential law far away from the mid-plane of the disc.

Note that the scale length estimates of the outer segment are very sensitive to the quality of background subtraction.
Moreover the truncation law of inner disc is unknown and may depend on the ram pressure stripping processes that in turn depend on the trajectory of the galaxy and on the disc orientation with respect to the incoming flow.
Therefore, one should not expect that the outer discs represent only a thick disc, and the radial scale there to correspond to the true value of thick disc radial scale.

\subsubsection{Vertical structure}

Next, we investigated the vertical structure of the edge-on discs in the studied galaxies.
We pursued two aims: i) to check whether vertical structure of the discs changes with radius particularly around $R_{\rm knee}$ and ii) to demonstrate how the contribution of the embedded thinnest disc component changes with $R$.
We assessed the parameters of the vertical profiles for a given galactocentric distance $R$ by using multi-component model \citep{Spitzer1942}:

\begin{equation}
I_z(R,z) = \sum_i \mu_i(R) \sech^{2}\left(\frac{z}{z_{0i}(R)}\right),
\label{eq_sech2}
\end{equation}

where $\mu_i$ and $z_{0i}$ denote the mid-plane surface brightness and the scale-height of each disc component.

We started to construct a two component model.
Unfortunately, such a model does not fit well extended wings of the vertical profiles.
This problem was recently addressed in the paper by \citet{comeron2018} where the authors demonstrated that PSF effects could be partly responsible for the extended wings at least in the \textit{Spitzer} data.
We utilized Tiny Tim PSF modeling tool \citep{TinyTim} and tested the PSF effects in the similar manner as in \citet{comeron2018}.
We concluded that the extended wings of vertical profiles in the studied galaxies cannot be described by PSF effects in the HST data\footnote{We also tested the effects of increasing thickness with radius and deviation from the precise edge-on orientation by means of integration of three-dimensional model of galaxy light along line-of-sight.
None of the effects make it possible to describe the extended wings in the profiles.
A detailed description of our experiments will be given in forthcoming paper.}.
This could  partly be due to the fact that we used modeled TinyTim PSF which has limited extension $R_{\rm PSF}=30$~arcsec.
Anyway, this has motivated us to apply a three-component model consisted of thin, mid, and thick components without additional PSF treatment.
Note, that we do not focus on the physical interpretation of each disc component because we have only long-slit data along the mid-plane of the studied galaxies.
This issue could be studied using long-slit data taken in an orientation perpendicular to the mid-plane or with IFU data.
Thus, the recently announced Fornax 3D project \citep{fornax3d}, including MUSE observations of our studied galaxies, would be particular useful for this aim.

We computed vertical profiles as a median of the galaxy cuts in the small radial bins of $\pm1$~arcsec size covering radial distances from $R_{\rm b}$ to $R=80...90$~arcsec where surface brightness drops down to $23-24$~mag arcsec$^{-2}$ at the mid-plane.
Since vertical profile decompositions into three components are not unambiguous and are affected by degeneracies between the model parameters we applied the following trick.
Firstly we fitted vertical profile which is closest to the $R_{\rm b}$ by using hand-tuned initial parameter guesses. Then, for the next profile we used the output parameters from the previous step as an initial guess and allowed that the scale heights of all components $z_{0i}$ to vary in some range around the initial guess.
The allowed scale height range was calculated from the condition that the gradient of vertical scale should be less than 0.5 arcsec per arcsec in radial distance.
We decomposed the vertical profiles by means of non-linear minimization method using \textsc{lmfit} package \citep{lmfit}.

\begin{figure*}
\centerline{
    \includegraphics[width=\textwidth,trim=0.25cm 0 0.2cm 0,clip]{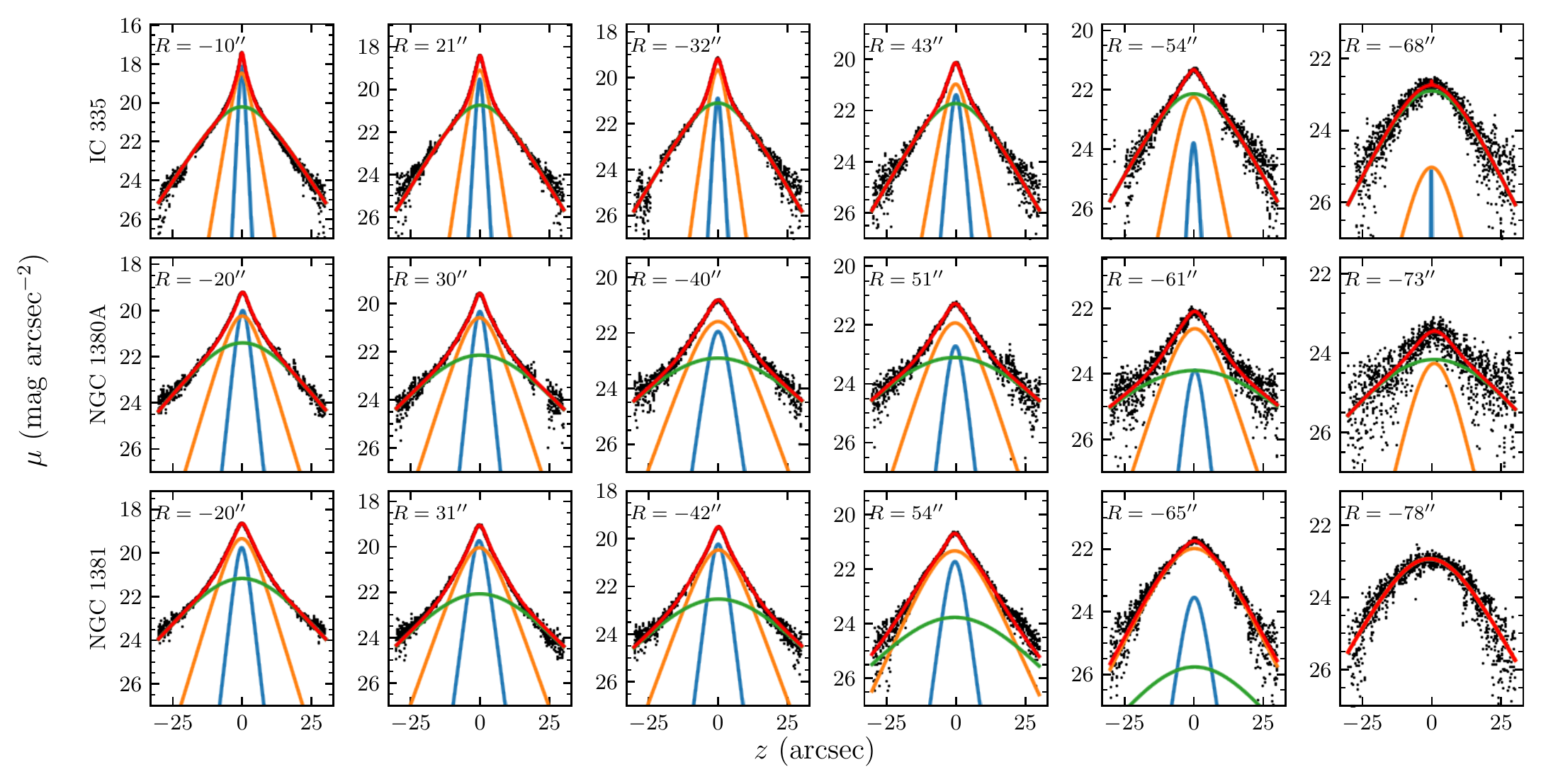}
}
\caption{Examples of decompositions of the vertical surface brightness profiles by using a three-component model.
Each row demonstrates vertical decomposition for a given galaxy.
Vertical profiles are sorted by $|R|$ from left to right.
Black dots show the image data in ACS/WFX F850LP filter; red lines are best-fit models; blue, orange, and green lines correspond to thin, mid, and thick component of our model.}
\label{phot_vert_decomp_profiles}
\end{figure*}

A scale height of the different components as a function of $R$ is shown in blue, orange, and green lines in Fig.~\ref{phot_decomp} (middle row) while the surface brightness along the mid-plane for the same model components and the observed profiles (black symbols) are presented in the bottom row panel.
Fig.~\ref{phot_vert_decomp_profiles} demonstrates a few examples of vertical profiles with overplotted best-fit models and separate components at different radial distances.
From our analysis we concluded that i) the contribution of the thin disc component (that is associated with the thinnest and probably with some additional mixture of the mid component) significantly decreases around $R_{\rm knee}$, and the most vertically extended components (mid and thick ones) dominate in the outer disc; ii) all disc components demonstrate moderate or significant growth of their thickness with the radius (except the regions where the component contribution is negligible).

In the work by \citet{comeron2018} there is photometric analysis of our galaxies based on \textit{Spitzer} data.
It is difficult to directly compare their results with ours because for each galaxy they considered the mean scale height values in four segments ($0.2r_{25}<|r|<0.5r_{25}$ and $0.5r_{25}<|r|<0.8r_{25}$) implying that scale height is constant within each segment.
Moreover, some segments can include our $R_{\rm knee}$.
For the internal segments ($0.2r_{25}<|r|<0.5r_{25}$), they obtained 1.4~arcsec and 7.1~arcsec for the scale heights of the thin and thick disc components of IC~335 ($r_{25}=78.9$~arcsec); 2.1~arcsec and 10.5~arcsec~-- for NGC~1380A ($r_{25}=82.6$~arcsec); 2.2~arcsec and 10.7~arcsec~-- for NGC~1381 ($r_{25}=75.4$~arcsec).
For the external segments ($0.5r_{25}<|r|<0.8r_{25}$) in most cases they did not get good fits.
Note that they computed the scale-heights by parametrising with an exponential function external parts of the brightness surface profiles of thin and thick disc components computed by solving the hydrostatic equilibrium equation.
Therefore their scale height estimations are a factor of 2 lower than those derived by using $\sech^2$-like parametrisation in our paper.
Nevertheless, taking into account such difference and comparing both approaches we found that our estimation of the scale height for the thinnest components is $2-4$ times lower than their values for the thin discs.
Our mid-scale components have in general compatible values with their thin disc measurements, while our thick components have slightly smaller scale-heights than their thick discs.
Significant differences in the estimates for the thinnest components could be naturally explained by difference in the spatial resolutions of \textit{Spitzer} and HST data despite the fact that the PSF effects have been taken into account by \citet{comeron2018}.

\subsubsection{Isophote analysis}

\begin{figure}
\centerline{
    \includegraphics[width=0.48\textwidth]{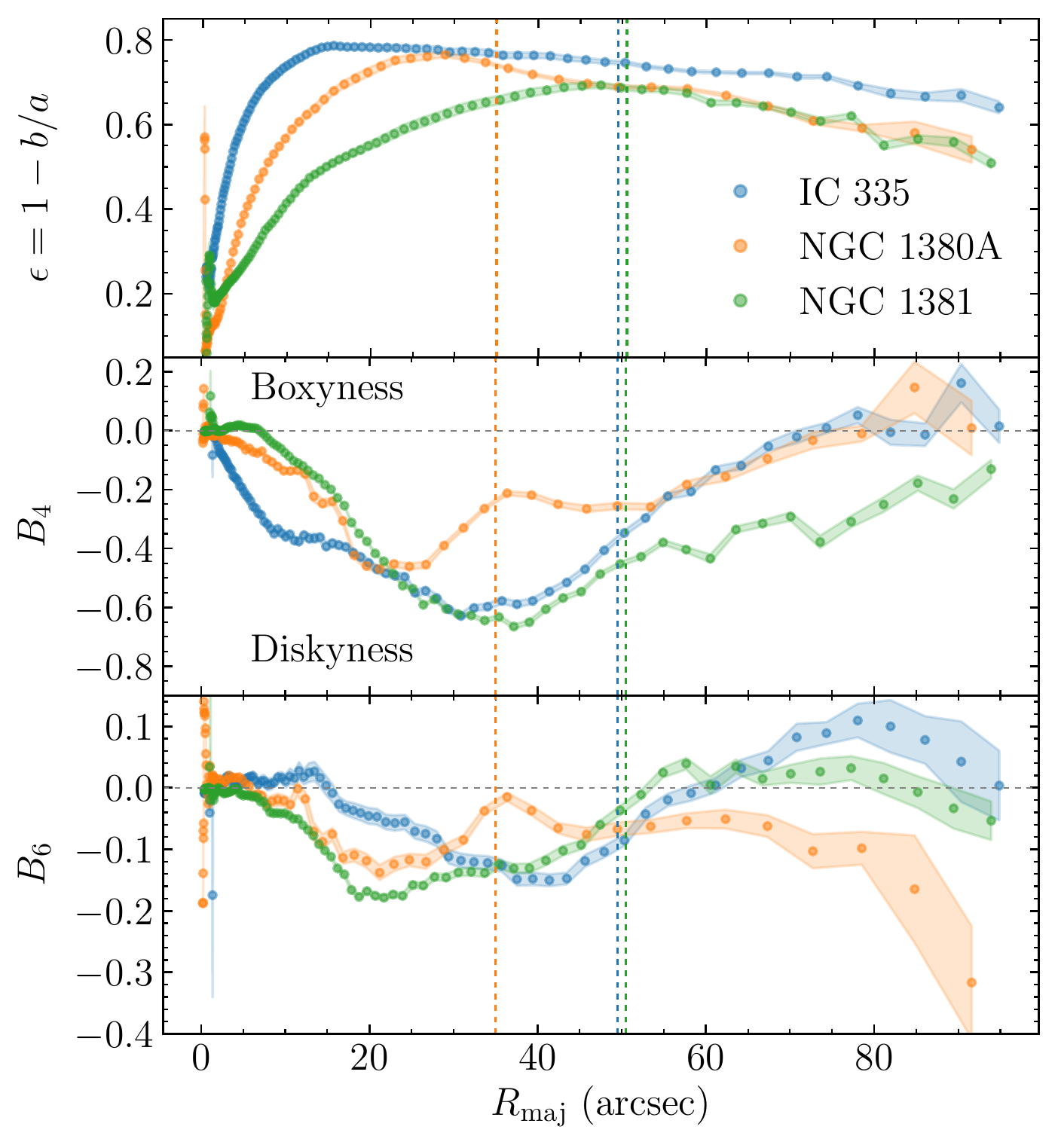}}
\caption{Results of the isophote analysis by means of {\sc Isofit} formalism \citep{Ciambur_ISOFIT_2015}.
The coefficients of the Fourier harmonics $B_{\rm 4}$, $B_{\rm 6}$ describe symmetrical deviations from ellipticity and are shown in two bottom panels.
The $B_{\rm 4}$ coefficients are particularly useful at capturing the boxy/disky shape of isophotes.
Shaded areas correspond to the parameter uncertainties.
Vertical dashed lines show $R_{\rm knee}$ values.}
\label{fig_ell}
\end{figure}

We applied the fitting formalism {\sc Isofit} recently developed by \citet{Ciambur_ISOFIT_2015} to the images.
This formalism provides an appropriate description of deviations from ellipticity and, therefore, is useful for isophote analysis in edge-on galaxies \cite[see examples in][]{Ciambur2016_edgeons}.
The resulting ellipticity and the $B_{\rm 4}$, $B_{\rm 6}$ coefficients of the Fourier harmonics are presented in Fig.~\ref{fig_ell}.
We used eight harmonics in the {\sc Isofit} tool.
Fig.~\ref{fig_ell} clearly demonstrates that all three objects have disky isophotes in their inner parts.
The parameter $B_4$ (negative $B_4$ values indicate diskyness) increases beyond the knee radius for the three galaxies and shows that the outer disc, although disky, is less disky than the inner disc.
Note that numerical simulations indicate that thick disc originated by minor mergers has boxy isophotes with respect to the main thin disc \citep{Villalobos2008,Purcell2010}.

\subsubsection{Photometry analysis summary}

To sum up, our photometric analysis supports the fact that the considered galaxies have more than one disc component, since i) their vertical profiles have a complex structure and are not fitted by a single or even two disc components within $R<R_{\rm knee}$; ii) the scale-lengths grow with $z$ again indicating a complex vertical disc structure and iii) isophotes change their shape sharply, reducing the diskyness beyond $R_{\rm knee}$.

We interpret that the thick disc components increase their contribution to the total light around $R_{\rm knee}$ and dominate the disc peripheries. It results in significant variations of the stellar population properties as a function of the radius.

This is in good agreement with the recent paper by \citet{comeron2018} where thick discs of edge-on galaxies are studied in the S$^4$G Survey.
The authors have demonstrated that thick discs are nearly ubiquitous, and their contribution to the surface brightness in the mid-plane can increase as radius grows (see their Fig.~22 and similar figures in their appendices for the galaxies studied here).

\section{Discussion}
\label{sec:discussion}

We have studied three edge-on galaxies (IC\,335, NGC\,1380A and NGC\,1381) belonging to the Fornax cluster.
Our study reveals important information for the stellar disc formation theory complementing the investigation of these galaxies by other authors \citep[for instance,][]{Bedregal2008stpop,Spolaor2010stpop, koleva} .

\begin{figure}
\centerline{
    \includegraphics[width=0.48\textwidth]{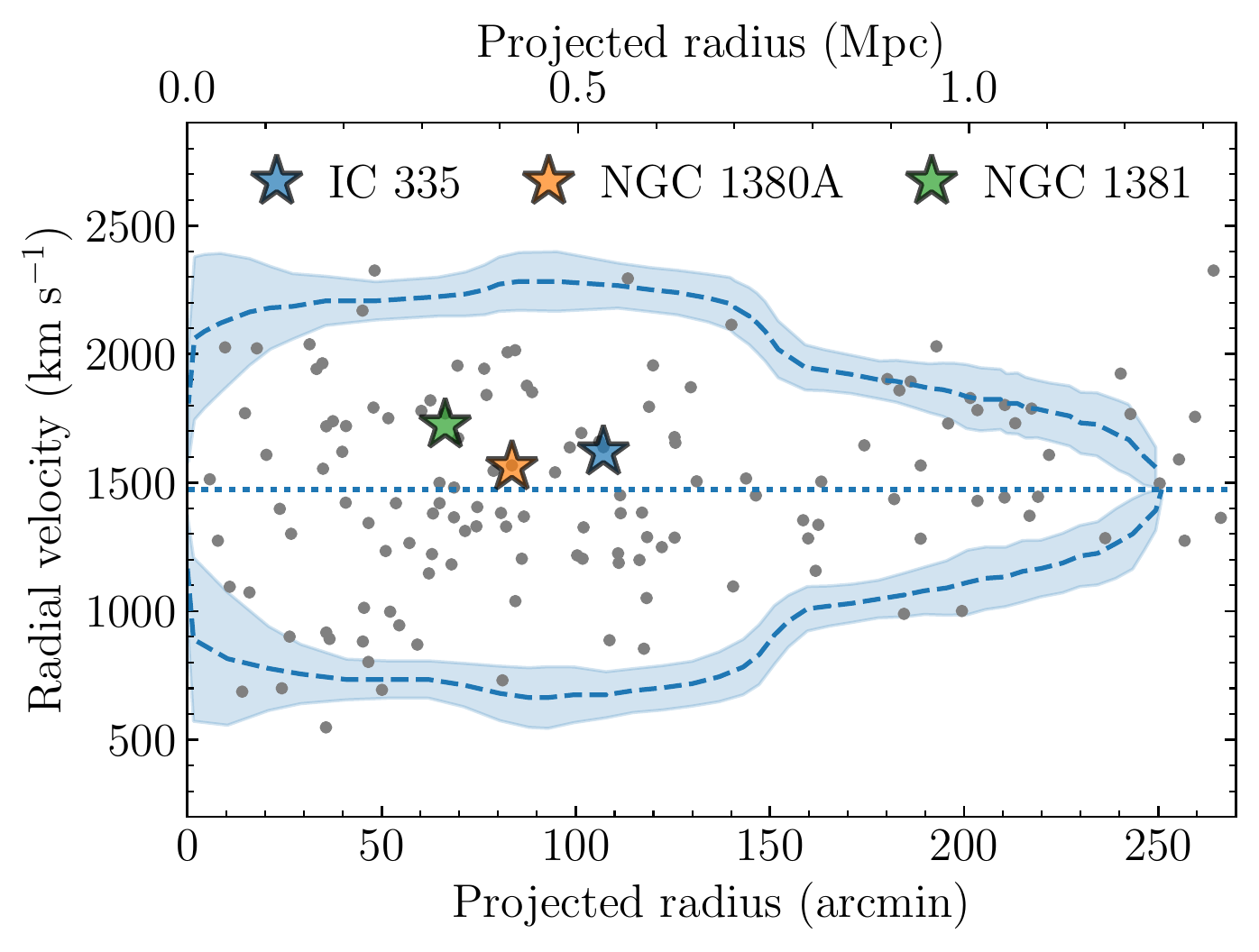}
}
\caption{Projected distances and radial velocities for Fornax cluster members (gray points). The blue dashed line shows a caustic curve calculated by \citet{Drinkwater2001}. It roughly corresponds to the escape velocity for a galaxy at a given distance from the cluster center. This diagram shows that our studied galaxies are dynamically bound to the main core of the Fornax cluster.}
\label{fig_caustic}
\end{figure}

The galaxies under consideration naturally fit into a two-phase model of galaxy assembly.
First, thick stellar discs formed rapidly at high redshifts in dense turbulent unstable gas-rich discs \citep{ElmegreenElmegreen2006,Bournaud2009ApJ,Comeron2014concurentGrowth,Elmegreens2017}.
After that the thin disc components grew for a long time from gas freshly accreted through cosmological filaments \citep{Sancisi2008A&ARv,Combes2014}, minor gas-rich mergers \citep{Robertson2006ApJ,Sancisi2008A&ARv}, by accretion of cooled left-over gas \citep{Burkert1992} or coronal gas cooled by the fountain mechanism \citep{Fraternali2009IAUS,Fraternali2013ApJ}.

Our galaxies belong to the Fornax cluster which has a complex structure and where mass assembly processes are still going on \citep{Drinkwater2001,iodice2017,Spiniello2018pn}.
Our galaxies are located near the main cluster core associated with NGC\,1399 where the majority of galaxies are of early type \citep{Ferguson1989}.
Fig.~\ref{fig_caustic} demonstrates that these galaxies are strongly dynamically bound to the main core of the Fornax cluster.
It is natural to assume that they could have experienced dense-environment effects \citep{BosellGavazzii2006PASP} in the past such as ram pressure stripping \citep{GunnGott1972, Abadi1999MNRAS, Quilis2000}, tidal interactions with the cluster gravitational potential and high-speed galaxy-galaxy encounters \citep{Moore1996Natur, Moore1998ApJ}, which could affect the galaxy evolutionary phase while the thin disc grew.

All our galaxies demonstrate a significant increase of the SSP equivalent age and decline of the stellar metallicity towards the galaxy peripheries (see Fig.~\ref{fig_specfit}), where the thick disc components dominate.
Recently \citet{kasparova} studied the edge-on galaxy NGC\,4710 belonging to the Virgo cluster, which has a similar behavior.
They proposed that the H\i\ gas layer has been stripped by ram pressure and as a result its thin disc is ``unfinished'' and we can observe a thick disc stellar population at the outskirts of NGC\,4710.
The same scenario takes place for the galaxies studied in this paper.
Note that ram pressure can effectively remove the gas starting from some particular radius where the ram pressure overcomes the gravitational attraction of the disc \citep{BosellGavazzii2006PASP}.
We suggest that $R_{\rm knee}$ could be this radius.

Intermediate-age stellar populations in the inner discs and the absence of any emission lines in the spectra point out the quenching of active star formation some time ago.
The most obvious explanation for this is a process of starvation \citep{Larson1980ApJ,Bekki2002ApJ,Bekki2009MNRAS,Zinger2018MNRAS}, which consists in removing the extended gas reservoir from a galaxy halo.
This results in the quenching of further star formation activity after a few Gyrs.
Due to the fact that SSP-equivalent ages are strongly biased towards the age of the younger population \citep{SerraTrager2007}, we can consider the stellar ages in the thin-disc dominated regions as a time stamp for the quenching of active star formation.
Hence, the  star formation in the main discs of IC\,335 and NGC\,1380A stopped approximately $4-5$~Gyr ago due to the dense cluster environment.
The inner disc of NGC\,1381 demonstrates an older age and a higher $\alpha$-element enhancement of its stellar population ($T_{\rm SSP}=7.3$~Gyr, [Mg/Fe$]=0.2$) with respect to other galaxies.
This indicates that the star formation in this galaxy was rapidly quenched in earlier epoch and can be explained if it entered earlier into the dense environment and if it was already in place $7-8$~Gyr ago ($z\approx1$).
Another feature of this object is the presence of a prominent bulge, therefore, bulge-driven processes, for instance, morphological quenching \citep{Martig2009} or active galactic nucleus feedback \citep{DiMatteo2005Natur,Croton2006MNRAS}, could be alternatively responsible for the early star formation quenching, without any relation to the cluster environment.

Recent investigations of lenticular galaxies in different environments \citep{Silchenko2012,Katkov2014MNRAS,Katkov15} have led to the scenario of general evolution of disc galaxies formulated by \citet{Silchenko2012}.
The main idea is that lenticulars are primordial disc galaxies which formed at high redshift ($z=2-3$) as a thick disc component \citep{ElmegreenElmegreen2006,Bournaud2009ApJ}.
The further fate of the galaxy strongly depends on the mode of gas accretion into their discs.
If there is persistent gas accretion and dynamical gas cooling, spiral arms can develop and star formation reignites: the galaxy is transformed into a typical spiral.
In absence of a gas-accretion source, which most commonly happens in dense environments, the galaxies preserve their lenticular morphology during all their life; therefore S0s are the dominant galaxy population in galaxy clusters at $z=0$.
Similar idea has also been suggested by \citet{Comeron2016AA_eso243-49} for the ESO~243-49 evolution.

In the galaxies investigated in this paper we have found imprints of all galaxy formation stages discussed above: the primordial formation of the thick discs in the outermost regions and the subsequent development of a thin disc component in their internal parts that has been stopped by environmental effects within the Fornax cluster.


\section{Summary}
\label{sec:summary}

In this paper we have performed a detailed study of three edge-on galaxies (IC\,335, NGC\,1380A, NGC\,1381) that belong to the Fornax cluster.
We explored publicly available photometrical HST data as well as new deep spectroscopic observational data obtained at the 10-m SALT telescope.

We have demonstrated that the long-slit spectra obtained with the RSS spectrograph have sufficient scattered light to bias the measurements of the stellar population properties in the outer parts of galactic discs.
We have developed a framework to take into account the scattered light which can be used for any kind of long-slit data analysis.

The stellar population properties of the outer disc regions in all three galaxies demonstrate a significantly older ages and lower metallicity than the inner ones.
Combining these data with a photometric analysis we have concluded that the changes of the stellar population properties are caused by an increase of the thick-disc contribution in the outermost galactic regions.
We interpreted this in the frame of a two-phase process of disc galaxy assembly where the thick disc component formed at high redshift while the thin disc developed later from the gas accreted from outside.
We suggest that the star formation in the outer disc has been quenched due to ram pressure stripping at the beginning of the cluster assembly while the rest of star formation in the discs was gradually extinguished by starvation.

\section*{Acknowledgements}
We are very grateful to the referee S{\'e}bastien Comer{\'o}n for comments and suggestions that improved this manuscript.
We also thank Prof. Anatoly Zasov and Dr. Igor Chilingarian for fruitful discussions.
The spectroscopical observations reported in this paper were obtained with the Southern African Large Telescope (SALT),
under programmes \mbox{2014-2-MLT-001} and \mbox{2015-2-MLT-002} (PI: Alexei Kniazev).
AYK acknowledges the support from the National Research Foundation (NRF) of South Africa.
IYK, AVK are grateful to the Russian Science Foundation grant 17-72-20119 which supported the photometrical analysis as well as the manuscript preparation.
IYK is also thankful to the RFBR grant number 16-02-00649.
The authors acknowledge partial support from the M.V.~Lomonosov Moscow State University Program of Development.
Based on observations made with the NASA/ESA Hubble Space Telescope, and obtained from the Hubble Legacy Archive, which is a collaboration between the Space Telescope Science Institute (STScI/NASA), the Space Telescope European Coordinating Facility (ST-ECF/ESA) and the Canadian Astronomy Data Centre (CADC/NRC/CSA).
This research made use of Astropy, a community-developed core Python package for Astronomy \citep{astropy2018}; The Atlassian JIRA issue tracking system and Bitbucket source code hosting service.




\bibliographystyle{mnras}
\bibliography{biblib} 







\bsp	
\label{lastpage}
\end{document}

%% file: tbl_stpop_binned.tex
\begin{table}
\begin{center}
\caption{Stellar population parameters in binned spectra.}
\begin{tabular}{lcc}
\hline
\hline
\multirow{2}{*}{Parameter}& \multicolumn{2}{c}{Disc segments} \\
\cline{2-3}
 & Inner disc & Outer disc \\

 \hline
 \multicolumn{3}{c}{IC 335} \\
 \hline
Binned regions, arcsec & 10\dots47 & 53\dots64 \\
T$_{\rm SSP}$, Gyr & $4.95^{+0.11}_{-0.11}$ & $11.07^{+2.24}_{-1.86}$  \\\relax
[Fe/H]$_{\rm SSP}$, dex & $0.10 \pm 0.01$ & $-0.59 \pm 0.05$  \\\relax
[Mg/Fe]$_{\rm SSP}$, dex & $0.07 \pm 0.02$ & $0.11 \pm 0.11$  \\

 \hline
 \multicolumn{3}{c}{NGC 1380A} \\
 \hline
Binned regions, arcsec & 12\dots32 & 38\dots60 \\
T$_{\rm SSP}$, Gyr & $3.89^{+0.18}_{-0.17}$ & $6.45^{+0.89}_{-0.78}$  \\\relax
[Fe/H]$_{\rm SSP}$, dex & $-0.08 \pm 0.01$ & $-0.33 \pm 0.05$  \\\relax
[Mg/Fe]$_{\rm SSP}$, dex & $0.09 \pm 0.04$ & $0.21 \pm 0.14$  \\

 \hline
 \multicolumn{3}{c}{NGC 1381} \\
 \hline
Binned regions, arcsec & 20\dots47 & 53\dots78 \\
T$_{\rm SSP}$, Gyr & $7.26^{+0.18}_{-0.18}$ & $11.73^{+1.88}_{-1.62}$  \\\relax
[Fe/H]$_{\rm SSP}$, dex & $-0.04 \pm 0.01$ & $-0.58 \pm 0.04$  \\\relax
[Mg/Fe]$_{\rm SSP}$, dex & $0.18 \pm 0.02$ & $0.21 \pm 0.08$  \\

\hline
\hline
\end{tabular}
\label{tab:stpop_binned}
\end{center}
\end{table}